\documentclass[%
 reprint,
 amsmath,amssymb,
 aps,
floatfix,
]{revtex4-2}

\usepackage{graphicx}
\usepackage{dcolumn}
\usepackage{bm}
\usepackage{amsfonts} 
\usepackage{amsmath}
\usepackage{booktabs}  
\usepackage{siunitx}  
\usepackage{color}

\begin{document}

\preprint{APS/123-QED}

\title{Comparing Repeated Gravitational-Wave Bursts Emitted by Cosmic Strings}

\author{Imene Belahcene,}
\email{imenemb@shao.ac.cn}
\affiliation{Shanghai Astronomical Observatory, Shanghai, 200030, China}%
\author{ Wen-Biao Han}%
\affiliation{Shanghai Astronomical Observatory, Shanghai, 200030, China \\
School of Fundamental Physics and Mathematical Sciences, Hangzhou Institute for Advanced Study, UCAS, Hangzhou 310024, China \\
School of Astronomy and Space Science, University of Chinese Academy of sciences, Beijing, 100049, China \\
Taiji Laboratory for Gravitational Wave Universe (Beijing/Hangzhou), University of Chinese Academy of Sciences, Beijing 100049, China}%

\date{\today}

\begin{abstract}

The principal energy loss mechanism in a Nambu-Goto cosmic string network involves loop production and the subsequent gravitational-wave emission. 
Recently, it has been shown that the loop oscillations produce repeated gravitational-wave bursts emitted at cusps.
The calculations are extended to estimate the number of burst repeaters, including kink emissions and kink-kink collisions.
Our findings indicate that despite the potentially large number of kinks, we anticipate observing a higher number of burst repeaters from cusps for both LIGO-Virgo-KAGRA and the Laser Interferometer Space Antenna.
We also conduct calculations using the second main loop distribution model in the current literature. 
We find this model predicts a notably higher number of repeaters, which provides a reason to include it in future data analysis.
These results offer insights into the potential observability of different loop features for future detectors, such as the space-based laser interferometers Taiji and TianQin.
\end{abstract}

\maketitle


\section{Introduction}\label{sec:1}

Cosmic strings can originate as one-dimensional topological defects from spontaneous symmetry breaking during phase transitions in the early Universe~\cite{Kibble:1976sj,Vilenkin:1984ib, Hindmarsh:1994re}. Topological defect formation is generic in grand unified theories~\cite{Jeannerot:2003qv}. Such phase transitions may have occurred at grand unification period, corresponding to an energy scale of about $10^{16}$~Gev. String theory-inspired cosmological models also predict the possibility of the formation of cosmic strings, which are sometimes referred to as "cosmic superstrings"~\cite{Witten:1985fp, Polchinski:2004ia, Sakellariadou:2010hr}. When two strings cross each other, they intercommute, wherein they exchange partners and form a closed loop. This process occurs with a probability referred to as $p$. The probability of intercommutation, as observed in both numerical simulations~\cite{Shellard:1987bv} and analytical models~\cite{Eto:2006db}, is typically assumed to be $p=1$. Nevertheless, in the case of superstrings, the probability of reconnection can be less than one due to the probabilistic nature of the interaction between fundamental strings. 

Cosmic string loops oscillate periodically in time, emitting gravitational waves. A loop of invariant length $\ell$, has period $T=\ell/2$ and decays in a lifetime $\tau=\ell/\gamma_{d}$ with $\gamma_{d}=\Gamma_d G\mu$ where $G\mu (c = 1)$ is the dimensionless string tension,  with G the Newton’s constant and $\mu$ the string mass per unit length. Gravitational radiation can be generated by small-scale structures that form on cosmic string loops. Cusps are points where the string instantaneously reaches the speed of light~\cite{Damour:2000wa}. Kinks are created in pairs after each string intercommutation, this pair of kinks travel at the speed of light in opposite direction along the string. Cusps and kinks source bursts of beamed gravitational waves, while the collision of two kinks produce an isotropic burst. The number of cusps on a loop is usually of order unity, it can be an order of magnitude larger as shown in~\cite{Pazouli:2020qmj, Pazouli:2021orp}, but remains globally constrained below 10.  The number of kinks can be much larger than one. Numerical simulations of string loops that favor $\Gamma_{d} \sim 50$, provide a limit on $N_{k} = 200$~\cite{LIGOScientific:2021nrg, Allen:1991bk}. In the case of cosmic superstrings, their network consists of various types of strings, each possessing a different tension. When two such distinct strings intersect, they create Y-junctions~\cite{Binetruy:2009vt, Binetruy:2010bq, Binetruy:2010cc}, giving rise to a new string segment connecting the original ones at two vertices. As a kink propagates through a Y-junction, it undergoes exponential proliferation, leading to an uncertain constraint on the maximum number of kinks at the moment. 

The detection of gravitational-wave emissions from cosmic string and superstring loops holds immense potential in advancing our understanding of the Universe. The superposition of gravitational-wave bursts gives rise to a stationary and nearly Gaussian stochastic background which can be probed over a large range of frequencies by different observations. In the absence of detection, the LIGO-Virgo-KAGRA (LVK) collaboration~\cite{LIGOScientific:2021nrg} and the Pulsar Timing Array (PTA) collaborations~\cite{EPTA:2023hof}, impose severe constraints on the dimensionless string tension $G\mu$.

Another promising way of detecting the presence of cosmic strings is through the single gravitational wave emission from loops. Searches for cosmic string signals with well-known waveforms~\cite{Damour:2000wa} are being carried out by the LVK collaboration using a matched-filter technique. In this paper we extend the work of~\cite{Auclair:2023mhe} which proposed the idea that loop oscillation produces burst repetitions. It focuses on the case of cusp emission, calculating the typical rate and period of repeating bursts that could be detected by LVK and LISA detectors. 

In Sec.~\ref{sec:2}, the gravitational-wave emission from cosmic string loops is introduced using a loop distribution model partly derived from numerical simulations of a Nambu-Goto string network, denoted as $M = 1$~\cite{Blanco-Pillado:2013qja}.
In Sec.~\ref{sec:3}, we expand upon the work presented in~\cite{Auclair:2023mhe} by including emissions from kinks and the collision of two kinks.
In addition, in Sec.~\ref{sec:4} we calculate the number of burst repeaters using an alternative loop distribution model, denoted as $M = 2$ and detailed in~\cite{Lorenz:2010sm}, which projects a larger loop population.
Sec.~\ref{sec:5} is dedicated to the discussion.

\section{Gravitational waves from cosmic string loops}\label{sec:2}

Considering the Nambu-Goto dynamics as the governing framework for cosmic strings, the emission of a burst signal or a stochastic gravitational-wave background predominantly arises from the oscillations of sub-horizon loops within the cosmic string network. Gravitational waves are produced by cusps, kinks and kink-kink collisions on cosmic string loops. The strain signal waveform is a power-law function of the frequency, $f$, derived in~\cite{Damour:2000wa}: 
\begin{equation}
    h(\ell,z,f)=A(\ell,z)f^{-q}, 
\end{equation}
where $\ell$ is the loop length, $z$ is the redshift and $q=\left \{ 4/3, 5/3, 2 \right \}$ for cusps, kinks, and kink-kink collisions. The signal amplitude is given by 
\begin{equation}
   A(l,z)=g_{1,q} \frac{G\mu \ell^{2-q}}{(1+z)^{(q -1)}r(z)}, 
\end{equation}
with $g_{1,q}=\left \{ 0.85, 0.29, 0.10 \right \}$ for cusps, kinks, and kink-kink collisions. This factor combines uncertainties in the waveform calculation. The comoving distance is denoted by $r(z)$, where we consider a $\Lambda$-CDM cosmological model, initially detailed in~\cite{Binetruy:2012ze}, and in the appendix. 
The beaming angle 
\begin{equation}
    \theta _{m}(\ell, z, f) = (g_{2} f (1+z) \ell)^{-1/3}, 
\end{equation}
represents the angular extent of the cone within which the majority of the gravitational-wave burst energy is concentrated. Here, $g_{2}=\sqrt{3}/4$ is a numerical factor. 

The crucial component for assessing the burst signal lies in characterizing the number of loops per unit loop size and per unit volume $\mathrm{d} ^{2} N/\mathrm{d}\ell \mathrm{d}V$.
The initial model developed, known as the one-scale model, was formulated by Kibble~\cite{Vilenkin:2000jqa}. This model describes the network evolution using a unique scale, the typical distance between the strings, in the scaling regime. The loops are formed at a fixed fraction of the horizon with the same relative size $\alpha$, given by $\ell = \alpha t$ at formation. It also assumes that loops do not self-intersect once formed. Although simple in nature, this model captures several characteristics of the network's evolution. Later, the model presented in \cite{Blanco-Pillado:2013qja} and referred here as $M = 1$ directly derives the loop production function for non-self-intersecting loops from simulations. Both of the aforementioned models produce a remarkably similar result, with a few factors at play. $M = 1$ assume, in the scaling regime, that the distribution of loops at time $t$ is :
\begin{align}
    t^{4}\frac{\mathrm{d} ^{2} N^{(1)}}{\mathrm{d}\ell \mathrm{d}V}_{r,r} &=\frac{0.18}{(\gamma + \gamma_{d})^{5/2}} \Theta (0.1 - \gamma) \\t^{4}\frac{\mathrm{d} ^{2} N^{(1)}}{\mathrm{d}\ell \mathrm{d}V}_{m,r} &=\frac{0.18}{(\gamma + \gamma_{d})^{5/2}} \left ( \frac{t_{eq}}{t} \right )^{1/2} \Theta (- \gamma +\beta(t)) \\
    t^{4}\frac{\mathrm{d} ^{2} N^{(1)}}{\mathrm{d}\ell \mathrm{d}V}_{m,m} &=\frac{0.27 - 0.45 \gamma^{0.31}}{(\gamma + \gamma_{d})^{2}}\Theta (0.18 - \gamma )\Theta (\gamma - \beta(t)),
\end{align}
which characterizes the distribution of loops during two distinct cosmic epochs: the radiation era (r,r) and the matter era (m,m). Specifically, it also considers the presence of residual loops from the radiation-dominated epoch, in matter era (m,r).
We denote by $t_{eq}$ the time of the radiation to matter transition. Here $\beta(t)$ is the relative length of the last loops formed in radiation era at a time $t_{eq}$.
This model predicts a reduced loop distribution compared to the one-scale model.
\\
\begin{figure}
\includegraphics[width=1.0\linewidth]{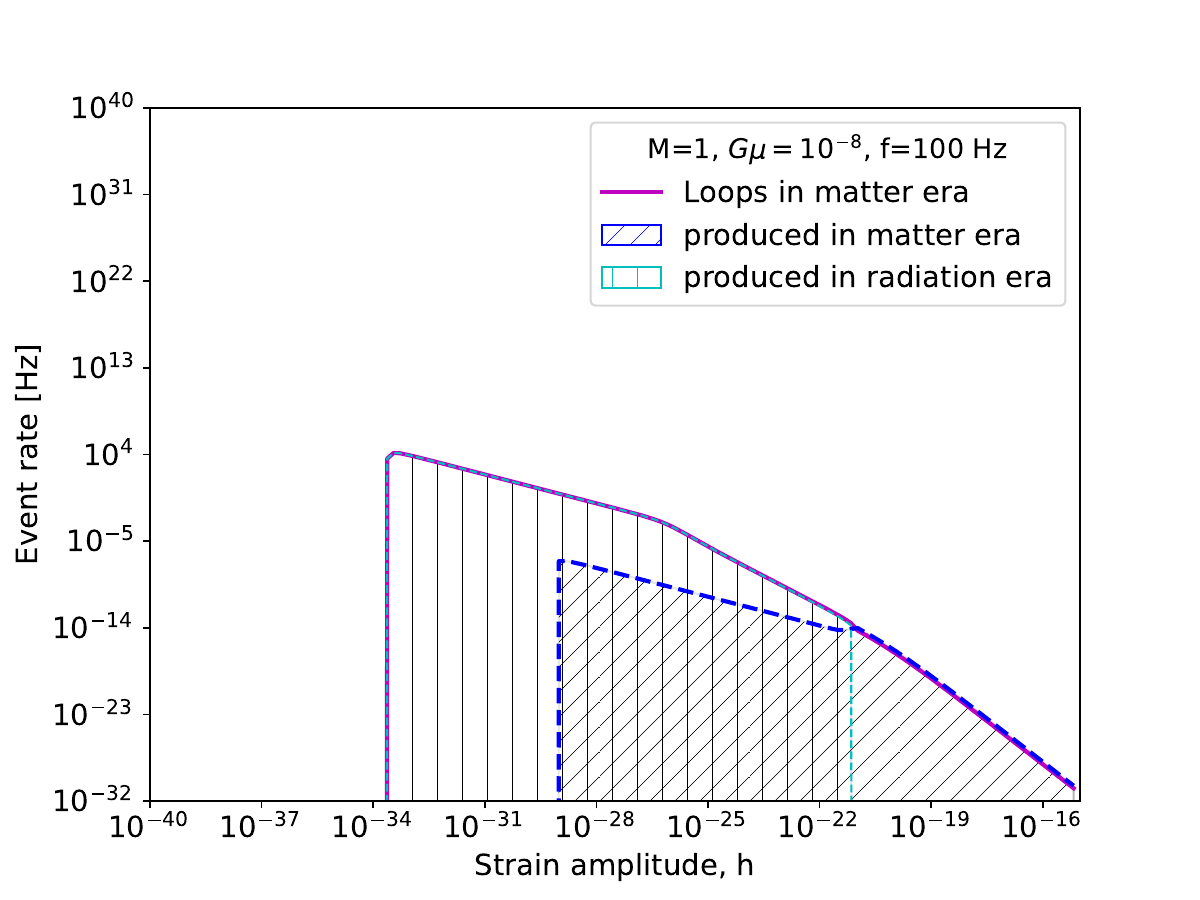}
\caption{\label{fig:rate_m2} Gravitational-wave event rate predicted by model M = 1 and averaged over the strain amplitude $h$. The string tension and the wave frequency are fixed to $G\mu = 10^{-8}$ and $f = 100$~Hz, respectively. For loops in the matter era, the effect of loops produced in the matter era is distinguished from the ones produced in the radiation era.}
\end{figure}
The rate of gravitational-wave bursts emitted at cusps or kinks we expect to detect from a proper volume $dV(z)$ at redshift $z$, derived in~\cite{Siemens:2006vk, LIGOScientific:2017ikf}, is given by 
\begin{equation}
\begin{split}
\frac{d^{2}R_{q}}{dzdh}(h,z,f) = & \frac{\phi_{V}(z)}{H_{0}^{3}(1+z)} \times \frac{2N_{q}}{(2-q)ht^{4}(z)}\\
& \times t^{4}\frac{\mathrm{d} ^{2} N}{\mathrm{d}\ell \mathrm{d}V}(\ell(h,z), t(z)) \\
& \times \Delta_q(hf^{q}, z, f), 
\end{split}
\end{equation}
where the cosmological parameters are defined in the appendix. 
Cusps emit gravitational-wave bursts in a cone of solid angle $d\Omega \sim \pi \theta_{m}^{2}$ and kinks into a fan-shaped set of directions in a solid angle $d\Omega \sim 2 \pi \theta_{m}$. Both emit in a narrow beams, meaning that only a small fraction of all cusps and kinks are oriented such that their gravitational-wave radiation can be observed on Earth. This fraction can be expressed as : 
\begin{equation}
    \Delta_q(\ell, z, f) = \left ( \frac{\theta_m (\ell, z , f)}{2} \right )^{3(2-q)}
\end{equation}
Figure~\ref{fig:rate_m2} shows that for a typical strength amplitude in the LVK detector for example of $h \sim 10^{-25}$, the burst search is more sensitive to the loops produced in radiation era, but still present in matter era. This result will be useful in the following. 
Note that the detected burst rate depends on the detector search efficiency to cosmic string gravitational-wave signals. 
The search efficiency is influenced by various factors, including the sensitivity of the interferometers, the noise level in the detectors, and the data analysis techniques used to search for and extract signals from the data. The search sensitivity is characterized by the minimum amplitude of a gravitational-wave signal that the detectors can reliably detect above the detector noise. For LVK~\cite{LIGOScientific:2021nrg} considering a characteristic frequency $f_{*}=20$~Hz we have $A_{*,c} \sim 2 \times 10^{-20} \, \text{s}^{-1/3}$ for cusps, $A_{*,k} \sim 6 \times 10^{-20} \, \text{s}^{-2/3}$ for kinks and $A_{*,kk} \sim 2 \times 10^{-19} \, \text{s}^{-1}$ for kink-kink collisions. For LISA~\cite{Auclair:2023brk} the data analysis was only conducted for cusps where the authors found $A_{*,c} \sim 3 \times 10^{-21} \, \text{s}^{-1/3}$ for a characteristic frequency $f_{*}=1$~mHz.

\section{Number of repeaters for kinks and collisions}\label{sec:3}

The general formula to express the number of loop small-scale structures observable is: 
\begin{equation}
   \frac{\mathrm{d}^2 N^{\cal{M}}_{q}}{\mathrm{d} \ell \mathrm{d} V} = \Delta_{q} \times N_{q} \times \frac{\mathrm{d}^2 N^{\cal{M}}}{\mathrm{d} \ell \mathrm{d} V}, 
    \label{eq:nb-features}
\end{equation}
where $\cal{M}$ refers to the loop distribution model and $N_{q}$ denotes the number of features on the loop. Specifically, we use $N_{c}$ for cusps, $N_{k}$ for kinks, and $N_{kk}$ for collisions.
To facilitate comparison, we maintain the numerical values chosen in the original article~\cite{Auclair:2023mhe}. Here, the time is fixed at $t=t_{0}=3 \times 10^{17}$ s, the age of the Universe. Additionally, we consider loops emitting at $z = z_{*} \ll 1$, redshift at which burst signals are expected to be detected. We first use the loop distribution predicted by model $M = 1$. By integrating Eq.(\ref{eq:nb-features}) with a change of variable and expressing the redshift as a function of the amplitude, we can determine the number of detectable kinks
\begin{equation}
\begin{split}
\frac{dN_{k}^{(1)}}{d \ell}(\ell) & = \int_{0}^{z^{*}} \frac{\partial^{2}  N^{(1)}_{k}}{\partial \ell \partial V} dV \\
& = N_{k} \times \frac{ 2 g_{1} ^{3}\pi }{3 g_{2}^{1/3}} \frac{C_{rad} t_{0}^{-3/2} \ell^{2/3}}{(\ell + \Gamma G\mu t_{0})^{5/2}}\sqrt{\frac{t_{eq}}{t_{0}}} \frac{(G\mu )^{3}}{A_{*k}^{3}f_{*}^{1/3}}, 
\end{split}
\end{equation}
and kink-kink collisions
\begin{equation}
\frac{dN_{kk}^{(1)}}{d \ell}(\ell) = N_{kk} \times \frac{ 4 g_{1} ^{3}\pi }{3} \frac{C_{rad} t_{0}^{-3/2} }{(\ell + \Gamma G\mu t_{0})^{5/2}}\sqrt{\frac{t_{eq}}{t_{0}}} \frac{(G\mu )^{3}}{A_{*kk}^{3}}.
\end{equation}
It should be noted that the amplitude of the signal emitted by the collision of two kinks remains independent of the loop size $\ell$, and the gravitational-wave emission is isotropic. Consequently, the number of observable collisions is unaffected by both the loop size and the detector frequency.
Considering the case where $\ell = \mathcal{O}(\text{yr})$ and $G\mu > 10^{-12}$, the denominator of these equations simplify, leading to a direct comparison with the results reported in the paper~\cite{Auclair:2023mhe} for the cusp case. We obtain the following expressions for the kinks
\begin{equation}
    \frac{dN_{k}^{(1)}}{d \ell}(\ell)  \approx N_{k} \times \frac{2 g_{1}^{3} \pi C_{rad} t_{0}^{-4} \ell^{2/3}}{3 g_{2}^{1/3} \Gamma^{5/2}} \sqrt{\frac{t_{eq}}{t_{0}}} \frac{\sqrt{G\mu}}{A_{*k}^{3}f_{*}^{1/3}}, 
\end{equation}
and for the kink-kink collisions
\begin{equation}
    \frac{dN_{kk}^{(1)}}{d \ell}(\ell)  \approx N_{kk} \times \frac{4 g_{1}^{3} \pi C_{rad} t_{0}^{-3/2} }{3  \Gamma^{5/2}} \sqrt{\frac{t_{eq}}{t_{0}}} \frac{\sqrt{G\mu}}{A_{*k}^{3}}.
\end{equation}
The burst repetition will enhance the signal-to-noise ratio, enabling more efficient detection. Substituting the minimum detectable amplitude $A_{}$ with the period-dependent amplitude $A_{} = A_{*}\sqrt{\ell/2T_{\text{obs}}}$~\cite{Auclair:2023mhe}, where $T_{\text{obs}}$ is the detector observation time, accounts for the gain in sensitivity due to burst repetition. Assuming that the sensitivity increase impacts the detectability of the repeaters, the number of detectable repeaters is higher. The expression for kinks in this context is 
\begin{equation}
      \frac{dN_{k}^{(1)}}{d \ell}(\ell)  \approx N_{k} \times \frac{2^{5/2} g_{1}^{3} \pi C_{rad} t_{0}^{-4} T_{0}^{3/2}}{3 g_{2}^{1/3} \Gamma^{5/2}  \ell^{5/6}} \sqrt{\frac{t_{eq}}{t_{0}}} \frac{\sqrt{G\mu}}{A_{*k}^{3}f_{*}^{1/3}}, 
\end{equation}
and for kink-kink collisions
\begin{equation}
      \frac{dN_{kk}^{(1)}}{d \ell}(\ell)  \approx N_{kk} \times \frac{2^{7/2} g_{1}^{3} \pi C_{rad} t_{0}^{-4} T_{0}^{3/2}}{3  \Gamma^{5/2} \ell^{3/2}} \sqrt{\frac{t_{eq}}{t_{0}}} \frac{\sqrt{G\mu}}{A_{*kk}^{3}}.
\end{equation}

\begin{figure}[htbp]
\includegraphics[width=1.0\linewidth]{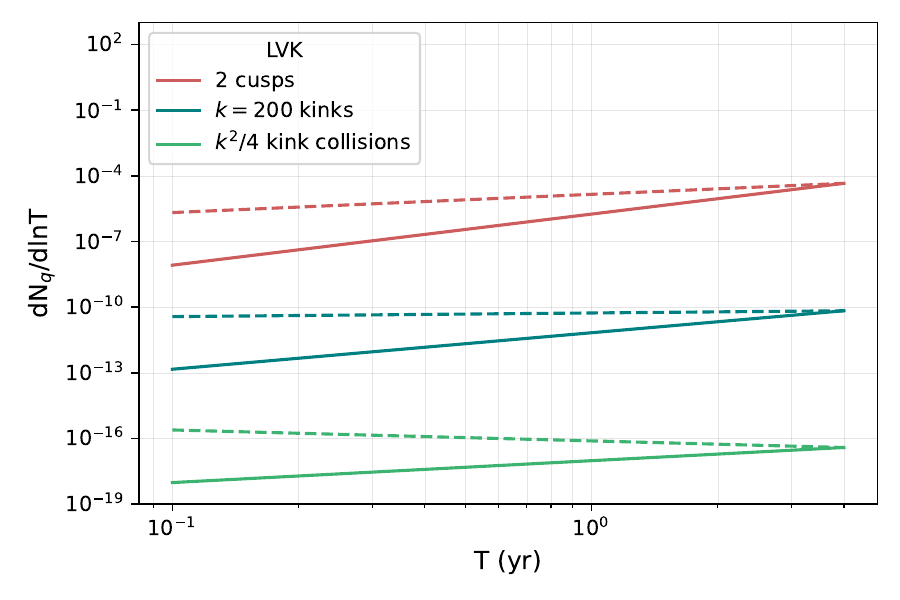}
\caption{\label{fig:epsart} Number of gravitational-wave burst repeaters per logarithmic bin of period T for LVK considering $G\mu = 10^{-10}$ and model $M = 1$. The solid line corresponds to the scenario where repetitions are ignored, while the dashed line illustrates the scenario considering the search sensitivity gain due to burst repetition with $T_{obs}=4$~yrs.}
\end{figure}

Figure~\ref{fig:epsart} compares the number of repeaters per logarithmic period bin in LVK for different types of emissions. The number of repeated burst is larger for cusps due to the strong dependence on the minimal detectable signal amplitude $A_{*}^{3}$. The lower sensitivity of the LVK burst search to kinks and collisions strongly affects the number of detectable repeaters. However, as the number of kinks increases, this is no longer trivial. We show that even with $N_{k} = 200$, the number of repeaters remains larger for cusps. Additionally, even in the context of strings created in string theory where $N_{k}$ could be higher, the contribution of cusps will dominate.

\section{Number of repeaters for $M = 2$}\label{sec:4}
Considering the model $M = 2$, an additional scale $\gamma_{c}$ is introduced, known as the backreaction scale. This scale accounts for the impact of gravitational-wave backreaction on the loop. Back-reaction reduces the number of small-scale wiggles, implying that strings have fewer opportunities to reconnect during each encounter, which hinders the formation of small loops. As a consequence, the loop distribution is defined across three distinct domains.
In the scaling regime, we disregard the influence of loops that were formed during the radiation era but persisted into the matter era. Our focus is exclusively on the loops formed during the matter era. The loop distribution is expressed as:

\begin{equation}
t^{4}\frac{\mathrm{d} ^{2} N^{(2)}}{\mathrm{d}\ell \mathrm{d}V}_{m,m}  =
    \begin{cases}
        \begin{aligned}
            &\frac{0.015}{(\gamma + \Gamma_{d}G\mu)^{3-2\chi_{m}}} \quad \text{if $\gamma_{d}<\gamma$}  \\
            &\frac{0.015(1-2\chi_m)}{(2-2\chi_m)\Gamma_{d}G\mu\gamma^{2-2\chi_m}} \quad \text{if $\gamma_{c}<\gamma<\gamma_{d}$}\\
            &\frac{0.015(1-2\chi_r)}{(2-2\chi_m)\Gamma_{d}G\mu\gamma_{c}^{2-2\chi_m}} \quad \text{if $\gamma<\gamma_{c}$} \\
        \end{aligned}
    \end{cases}
\label{eq:dist_m3}
\end{equation}
The numerical values for $\gamma_c$, $\chi_m$, and $\chi_r$ are obtained from the references~\cite{LIGOScientific:2021nrg, Lorenz:2010sm}.

\begin{figure}[htbp]
\includegraphics[width=1.0\linewidth]{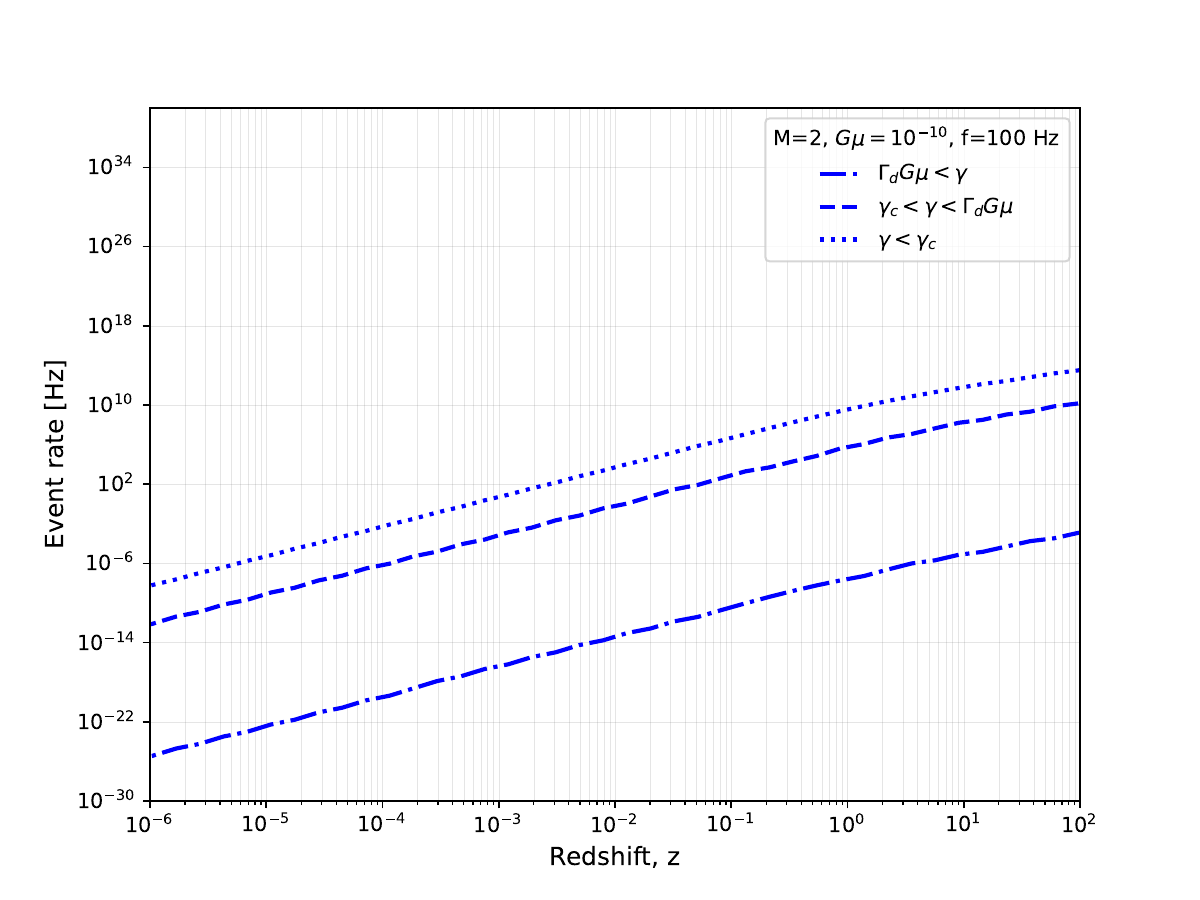}
\caption{\label{fig:rate_m3} Gravitational-wave event rate predicted by model $M = 2$ and averaged over the redshift $z$. The string tension and the wave frequency are fixed to $G\mu = 10^{-10}$ and $f = 100$~Hz, respectively.}
\end{figure}

\begin{figure}[htbp]
\includegraphics[width=1.0\linewidth]{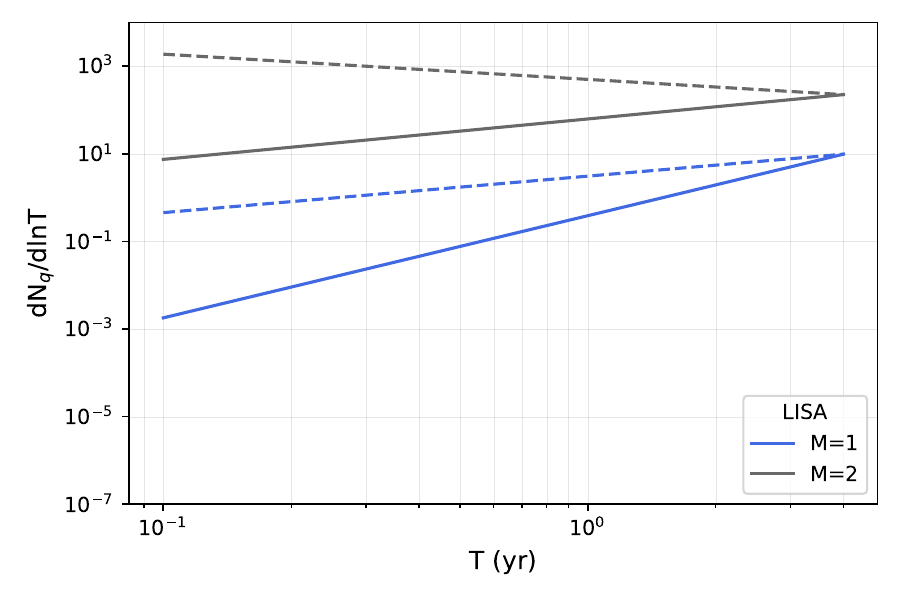}
\caption{\label{fig:m1m2} Number of gravitational-wave cusp burst repeaters per logarithmic bin of period $T$ in LISA for model $M = 2$ and $G\mu=10^{-10}$. The solid line corresponds to the scenario where repetitions are ignored, while the dashed line illustrates the scenario considering the search sensitivity gain due to repetition with $T_{obs}=4$~yrs.}
\end{figure}

Figure~\ref{fig:rate_m3} illustrates the burst rate for this model at low redshifts, displaying each contribution in the loop distribution individually. 
In this model, the abundance of small loops is proportional to the inverse power of the gravitational-wave backreaction scale $\gamma_{c}$, which is itself quite small. 
Consequently, this model predicts a considerably higher rate than $M = 1$.
In order to compare the number of burst repeaters with the first model, we ensure that we are working under the same conditions. By considering loop sizes $\ell = O(yr)$, we have $\gamma < \Gamma_d G\mu$ for $G \mu = 10^{-10}$. We work then with the intermediary loop distribution given in Eq.(\ref{eq:dist_m3}).
Using the same method as previously explained, we provide the direct expression for the number of detectable cusps without considering the sensitivity gain of the detector due to repetitions:
\begin{equation}
    \frac{\mathrm{d} N_{c}^{(2)}}{\mathrm{d} \ell} = \frac{2 g_{1}^{3} \pi}{3 g_{2}^{2/3} }\frac{0.015(1-2\chi_{m})t_{0}^{-(2+2\chi_{m})}}{(2-2 \chi_{m}) \Gamma_d \ell^{(2-6\chi_m)/3}}\frac{(G\mu)^2}{A_{*}^{3}f_{*}^{2/3}}, 
\end{equation}
and with
\begin{equation}
\frac{\mathrm{d} N_{c}^{(2)}}{\mathrm{d} \ell} = \frac{2^{5/2} g_{1}^{3} \pi}{3 g_{2}^{2/3} }\frac{0.015(1-2\chi_{m})t_{0}^{-(2+2\chi_{m})}T_{0}^{3/2}}{(2-2 \chi_{m}) \Gamma_d \ell^{(13/6-2\chi_m)}}\frac{(G\mu)^2}{A_{*}^{3}f_{*}^{2/3}}.
\end{equation}
Figure~\ref{fig:m1m2} displays the number of repeater for models $M = 1$ and $M = 2$ observed by LISA. Notably, LISA exhibits higher sensitivity to repeated bursts at $G\mu=10^{-10}$ for model $M = 2$. Table~\ref{tab:m1m2} presents the the number of repeaters produced by cusps, considering or not the gain in sensitiviy for LVK and LISA with $T=1$~yr.
\begin{table}[htbp]
\centering
\begin{tabular}{l@{\hspace{1.5em}}cc}
\toprule
\textbf{Type of Event} & \textbf{LVK} & \textbf{LISA} \\
\midrule
Single event & $3 \times 10^{-4}$ & \SI{63}{} \\
\addlinespace 
With repetitions & $2 \times 10^{-3}$ & \SI{502}{} \\
\bottomrule
\end{tabular}
\caption{Number of burst repeaters for cusps with period $T=1$~yr and $G\mu=10^{-10}$ for model $M = 2$ in LISA.}
\label{tab:m1m2}
\end{table}
Despite the current constraints, we present the result for $G\mu=10^{-10}$ for the $M = 2$ model. Currently, the results from LVK or PTA collaborations impose severe constraints on model $M = 2$, with $G\mu \leqslant 4 \times 10^{-15}$ being the most severe constraint set by the LVK. To derive these upper limits on the string tension $G\mu$, the gravitational-wave energy density spectrum predicted by a specific loop distribution model is compared with observational results.
The absence of a stochastic gravitational-wave background detection is then used to compute an upper limit on $G\mu$. This spectrum is derived for the case of Nambu-Goto strings, which evolve in the standard cosmological background. Considering more exotic cases can lead to a significantly different spectrum and potentially ease current constraints. One possible approach to circumvent the current constraint on $G\mu$ is through the presence of a metastable cosmic string network~\cite{Leblond:2009fq}. The metastability of the string network arises from its ability to undergo breaking via the formation of monopole-antimonopole pairs. This stops the intercommutation phenomenon, and thus the formation of loops. If we consider that the monopole reentrance occurs during the radiation era, the population of loops formed during the matter era no longer exists. This could allow  to modify the spectrum of the expected stochastic background.
Alternative scenarios in the history of the Universe  also predict a cutoff in the gravitational-wave spectra, which can potentially alleviate the current upper limits on $G\mu$ for $M = 2$~\cite{Gouttenoire:2019kij}. 

\section{Discussion}\label{sec:5}

Burst repeaters have been observed in various astronomical contexts, including events like gamma-ray bursts and fast radio bursts. A recent study has revealed the repetitive nature of gravitational-wave bursts emitted by oscillating cosmic string loops.
We further investigated this phenomenon by calculating the expected number of burst repeaters resulting from kinks or kink collisions using the loop distribution model referred to as $M = 1$. We found that, even with a large number of kinks (e.g., $N_k=200$), the number of burst repeaters is significantly higher for emission from cusps. 
\\
Additionally, we performed calculations using the model $M = 2$, which predicts a larger distribution of cosmic string loops. Our results indicate a significantly higher number of repeaters, with LISA showing interesting sensitivity in this scenario. However, it's worth noting that this model is heavily constrained by current observations. Nevertheless, these constraints can be relaxed when considering metastable cosmic strings which may introduce a cutoff in the gravitational-wave spectrum at a much higher frequency than the cutoff associated with stable string networks. Another way to alleviate these constraints is by exploring the evolution of the cosmic string network within a non-standard cosmological framework.
\\
It is important to note that the model $M = 1$ is considered as the most pessimistic among loop distribution models. In contrast, the one-scale model predicts a significantly more abundant loop distribution, with a factor of approximately 10, resulting in a higher number of burst repeaters. Furthermore, it's worth mentioning that when considering cosmic superstrings, the loop distribution can be scaled, for example, by a factor of $1/p$, where $p\leq 1$~\cite{Sakellariadou_2005, Avgoustidis_2006}.
\\
In LVK, matched filter burst search for cosmic strings face challenges from the presence of spurious non-Gaussian noises that can mimic the cosmic string signals, called blip glitches. The increasing sensitivity of LVK detectors has led to a growing number of these transient noises, significantly reducing the sensitivity of the cosmic string burst search. Short-duration glitches mimicking cosmic string burst signals could also appear in the LISA detector. The sequence of burst repeaters follows a predictable track in time-frequency space, unlike random glitches. This can help to distinguish a true cosmic string signal from a transient noise. Therefore, developing a template to search for burst repeaters for the $M = 2$ model in LVK, LISA and Taiji~\cite{Wu:2023rpn} remains interesting, despite current constraints. 
\\
\begin{acknowledgments}
This research was funded by the National Key R \& D Program of China, Grant No. 2021YFC2203002, NSFC (National Natural Science Foundation of China) No. 12173071; I.B. is supported by the Overseas Recruitment Programs; This work made use of the High-Performance Computing Resource in the Core Facility for Advanced Research Computing at Shanghai Astronomical Observatory.

\end{acknowledgments}

\appendix

\section{Standard cosmology}

We consider the case of a spatially flat Friedmann-Lemaitre-Robertson-Walker Universe. The Hubble rate at redshift $z$ is given by:
\begin{equation}
    H(z)=H_{0}\mathcal{H}(z), 
 \end{equation}
where
\begin{equation}
     \mathcal{H}(z)=\sqrt{\Omega_{\Lambda}+\Omega_{M}(1+z)^{3}+\Omega_{R}\mathcal{G}(z)(1+z)^{4}}, 
\end{equation}
$H_0$ represents the present value of the Hubble constant, and $\Omega_{i}$ denotes the present-day energy density, with the subscript $i$ being R for radiation, M for matter, and $\Lambda$ for the cosmological constant. We use the values given in~\cite{Planck:2015fie}.
In the standard model, entropy is assumed to be conserved and it is shared approximately among each of the relativistic species present.
The higher the temperature, the greater the number of species present.
As the Universe cools down, species become non-relativistic and release their entropy to the relativistic species that are still in thermal equilibrium.
In the radiation era, this is described by the function $\mathcal{G}(z)$ defined as~\cite{Bin_truy_2012}:
\begin{equation}
\begin{aligned}
    \mathcal{G}(z) \equiv \frac{g_{\ast}(z)g_{S}^{4/3}(0)}{g_{\ast}(0)g_{S}^{4/3}(z)} = 
    \begin{cases} 
        1 & \text{for } z < 10^{9} \\
        0.83 & \text{for } 10^{9} < z < 2 \times 10^{12} \\
        0.39 & \text{for } z > 2 \times 10^{12} 
    \end{cases}
\end{aligned}
\end{equation}
where $g_{\ast}(z)$ is the total effective number of degrees of freedom of all relativistic particles in thermal equilibrium at redshift $z$ and $g_{S}(z)$ is the effective number of entropic degrees of freedom.
The first equation represents the scenario when all Standard Model particles are relativistic. The second equation pertains to the phase after the quark-hadron transition (at $T\geqslant 200$~Mev), while the final one considers the stage after electron-positron annihilation and neutrino decoupling (at $T\geqslant 200$~kev).
The cosmic time can be expressed using the interpolation function $\varphi_t(z)$:
\begin{equation}
  t\rightarrow t(z)\equiv \frac{\varphi_{t}(z)}{H_{0}} \quad \textrm{with} \quad \varphi_{t}(z)\equiv \int_{z}^{+\infty}\frac{dz}{\mathcal{H}(z)(1+z)}.
\end{equation}
The proper distance $r(z)$ is expressed as:
\begin{equation}
  r\rightarrow r(z)\equiv \frac{\varphi_{r}(z)}{H_{0}} \quad \textrm{with} \quad \varphi_{r}(z)\equiv \int_{0}^{z}\frac{dz}{\mathcal{H}(z)}, 
\end{equation}
and the proper volume $V(z)$ as: 
\begin{equation}
  dV\rightarrow dV(z)\equiv \frac{\varphi_{V}(z)}{H_{0}^{3}}dz \quad \textrm{with} \quad \varphi_{V}(z)=\frac{4\pi \varphi_{r}^{2}}{(1+z)^{3}\mathcal{H}(z)} 
\end{equation}
The analytical calculation gives an asymptotic approximation of $\varphi_{r}(z)$ and $\varphi_{t}(z)$:
\begin{equation}
\begin{aligned}
    \varphi_{r}(z) &\approx \begin{cases}
        z & \textrm{for } z \ll 1\\
        3.2086 & \textrm{for } z \gg 1
    \end{cases} \\
    \varphi_{t}(z) &\approx \begin{cases}
        0.9566 & \textrm{for } z \ll 1\\
        \frac{1}{2z^{2}\sqrt{\Omega_{R}\mathcal{G}(z)}} & \textrm{for } z \gg 1
    \end{cases}
\end{aligned}
\end{equation}

\nocite{*}

\bibliography{ComparisonBurstRepeaters}

\end{document}